%% file: PrivPubGood.tex
\documentclass[12pt]{article}

	\newcommand{\blind}{0}
	\newcommand{\draft}{0} 
	
	\newcommand{\mytitle}{Suboptimal Provision of Privacy and Statistical Accuracy When They are Public Goods}
	
	\newcommand{\myversion}{\today  \if1\draft {\ \\ PRELIMINARY DRAFT -- PLEASE DO NOT CITE} \fi}
	\newcommand{\mythanks}{
	Abowd: U.S.\ Census Bureau and Department of Economics, Cornell University.
	Schmutte: University of Georgia and U.S.\ Census Bureau.
	Sexton: U.S.\ Census Bureau and Department of Economics, Cornell University.
	Vilhuber: Department of Economics, Cornell University and U.S.\ Census Bureau.
	We acknowledge the support of the Alfred P.\ Sloan Foundation Grant G-2015-13903 and NSF Grant SES-1131848. Abowd and Vilhuber acknowledge direct support from NSF Grants BCS-0941226 and TC-1012593. We are grateful for helpful comments from Larry Blume, David Card, Michael Castro, Cynthia Dwork, John Eltinge, Stephen Fienberg, Mark Kutzbach, Ron Jarmin, Dan Kifer, Ashwin Machanavajjhala, Frank McSherry, Gerome Miklau, Kobbi Nissim, Mallesh Pai, Jerry Reiter, Eric Slud, Adam Smith, Bruce Spencer, Sara Sullivan, Glen Weyl, and Nellie Zhao. Any opinions and conclusions expressed herein are those of the authors and do not  represent the views of the U.S. Census Bureau, NSF, or the Sloan Foundation. No confidential data were used in this paper.
	}

	\date{\myversion}
	\input{formats}

\begin{document}

	%JBES defined spacing
	\def\spacingset#1{\renewcommand{\baselinestretch}%
	{#1}\small\normalsize} \spacingset{1}

	\if0\blind
	{
	  \title{\bf \mytitle}
	  \author{John M.~Abowd, 
	     % U.S.\ Census Bureau and Department of Economics, Cornell University \\ 
	     % \href{mailto:john.maron.abowd@census.gov}{\texttt{john.maron.abowd@census.gov}}
	    % \and
	    Ian M.~Schmutte, 
	    % Department of Economics, University of Georgia\\ \href{mailto:schmutte@uga.edu}{\texttt{schmutte@uga.edu}}
	    % \and
	    William Sexton, 	    % U.S.\ Census Bureau and Department of Economics, Cornell University\\ \href{mailto:william.n.sexton@census.gov}{\texttt{william.n.sexton@census.gov}}
	    % \and
	    Lars Vilhuber\footnote{\mythanks}
	    % Department of Economics, Cornell University \\
	    % \href{mailto:lars.vilhuber@cornell.edu}{\texttt{lars.vilhuber@cornell.edu}}
	    }
	  \maketitle
	  \pagenumbering{gobble} %eliminate page number on front page
	} \fi

	\if1\blind
	{
	  \bigskip
	  \bigskip
	  \bigskip
	  \begin{center}
	    {\LARGE\bf \mytitle}
	\end{center}
	  \medskip
	} \fi

	% \bigskip
	% \clearpage
	% \pagenumbering{gobble} %eliminate page number on abstract page
	\begin{abstract}
	\input{abstract}
	\end{abstract}
	\noindent%
	\vfill

	\newpage
	\pagenumbering{arabic}
	\setcounter{page}{1}
	\setcounter{section}{0}
	\spacingset{1.45} % DON'T change the spacing!

% 	\section{Introduction}
% 	\label{sec:intro}
	\input{introduction}

	\section{Preliminary Concepts}
	\label{sec:preliminaries}
	\input{preliminaries}

	\section{Private Provision of Population Statistics}
	\label{sec:suboptimality}
	\input{suboptimality}

	\section{Conclusion}
	\label{sec:conclusion}
	\input{conclusion}

	% ------------------------------ Bibliography---------------------
	\bibliographystyle{\mybibstyle}
	\bibliography{library-excerpt,paper}
	%-----------------------------------------------------------------

	% ------------------------------ Appendices-----------------------
	\clearpage
	\appendix
	\section*{APPENDIX}
	\pagenumbering{arabic}\renewcommand{\thepage}{App.~\arabic{page}}
	\setcounter{section}{1}
	\label{sec:appendix}
	\input{appendix}

\end{document}

%% file: formats.tex
\renewcommand\footnotemark{}

\makeatletter
\def\blfootnote{\xdef\@thefnmark{}\@footnotetext}
\makeatother

\makeindex

\usepackage[printonlyused]{acronym}   % defines acronyms
\usepackage{natbib}    % for bibliography
\usepackage[named]{algo}
\newcommand{\mybibstyle}{dcu} % a harvard style
\usepackage{amssymb}  % AMS symbols
\usepackage{amsmath}  % AMS math
\usepackage{amsfonts}
\usepackage{appendix} % allows flexible appendix environments
\usepackage{graphicx} % for inclusion of EPS graphics
\usepackage{multicol} % Multicolumn formatting (in-line tables)
\usepackage{fullpage}  % enlarged margins
\usepackage{rotating}  % allows for rotated tables
\usepackage{epsfig}    % legacy package for inclusion of EPS
\usepackage{varioref}  % allows for references like ``Table x on page y''
\usepackage{float}     % ?
\usepackage{array}     % Modifications of tabular environment
\usepackage{setspace}
\usepackage{threeparttable} %Allows tables with automatic notes section
\usepackage{longtable} % allows for tables to wrap across pages.
\usepackage{supertabular} % allows for tables to wrap across pages.
\usepackage{fancyhdr}  % provides pagestyle fancy (page headers)
\usepackage[nottoc]{tocbibind} % provides for bibliography, List of tables, etc. to
\usepackage{color}
\usepackage{theorem}
\usepackage{layout}
\usepackage{makeidx}
\usepackage{acronym}
\usepackage{import}

\usepackage{tikz} %graphs
\usetikzlibrary{calc}
\usetikzlibrary{backgrounds} 
\usetikzlibrary{plotmarks}

\usepackage{xcolor} %font colors
\usepackage{multirow} %also for multi-columns
\usepackage{geometry} %manually laying out pages
\usepackage{threeparttable} %handy table set up
\usepackage{fullpage} %manually laying out pages
\usepackage{arydshln} %special lines in tables

\definecolor{myblue}{rgb}{0,.2,1}

\usepackage{hyperref}
	\hypersetup{%
	naturalnames=true,%
	bookmarksnumbered=true,%
	bookmarksopen=false,%
	plainpages=true,%
	colorlinks=true,%
	urlcolor=myblue,
	linkcolor=myblue,%
	filecolor=myblue,%
	citecolor=black,%
	pdfpagemode=UseOutlines%
	}%

\usepackage{palatino}

\bibpunct{(}{)}{;}{a}{}{;} %set if using natbib

\newcolumntype{M}{>{$}c<{$}}
\newcolumntype{R}{>{$}r<{$}}

\theoremstyle{plain}
\theorembodyfont{\normalfont}
\theoremheaderfont{\itshape}

\newtheorem{theorem}{Theorem}

\newtheorem{definition}{Definition}

\newenvironment{proof}[1][Proof]{\textbf{#1.} }{\ \rule{0.5em}{0.5em}}

\def\func#1{\mathop{\rm #1}\nolimits}%

%% file: abstract.tex
\noindent With vast databases at their disposal, private tech companies can compete with public statistical agencies to provide population statistics.
However, private companies face different incentives to provide high-quality statistics and to protect the privacy of the people whose data are used. When both privacy protection and statistical accuracy are public goods,  private providers tend to produce at least one suboptimally, but it is not clear which. We model a firm that publishes statistics under a guarantee of differential privacy. We prove that provision by the private firm results in inefficiently low data quality in this framework.

% \vspace*{2em}

% \noindent \textbf{Keywords:} Demand for public statistics; Technology for
% statistical agencies; Optimal data accuracy; Optimal confidentiality
% protection\newline

% \vspace*{1em} \noindent\textbf{JEL\ classification:} C40, C81, H41

%%% Local Variables:
%%% mode: latex
%%% TeX-master: abowd-schmutte-privacy
%%% End:

%% file: introduction.tex
Traditionally, statistical agencies have been charged with publishing summaries of data collected from the nation's citizens and businesses. Their data collection activities are expensive, and at risk of losing funding, despite an increasing demand for reliable data.\footnote{For evidence of increasing demand for Census data, see the discussion in \citet{Ruggles:DP:AEAPP:2019}.} 
In this environment, one option is to augment, or replace, public statistical programs with information collected by private companies. 
Indeed, companies that aggregate personal data---e.g. Facebook, Google, Apple,  Microsoft, Uber, Upwork---are under pressure to use their vast databases in the public interest. They are, often in collaboration with academic researchers, developing innovative data products like Google Trends \citep{choi2012predicting}, the Billion Prices Project \citep{Cavallo:BillionPrices:JEP:2016}, and the University of Michigan Social Media Job Loss Index \citep{Shapiro:SocialMediaLaborFlows:NBER2014}.
Clearly, the private sector is capable of producing innovative data products and could provide them competitively to the public.
% Note: could also include private data on firms - ADP - or retail or financial transactions - Nielsen, JCMB, Mint

Why are population statistics provided by public statistical agencies rather than private firms?
There are a number of potential difficulties, but in this paper we focus on inefficiencies in how private providers trade off data privacy and accuracy. Following the fundamental law of information recovery \citep{Dinur:2003:RIW:773153.773173}, increasing the accuracy of published statistical summaries necessarily results in a loss of privacy for the data owners. This means statistical agencies must perform a balancing act. Published statistics should be as accurate as possible without revealing too much information about any single individual or business. When the benefits of more accurate population statistics and privacy losses are shared by all citizens, we show that private provision will result in inefficiently low levels of data accuracy and inefficiently high levels of privacy protection.

To establish suboptimality of the private provision of population statistics, we model the problem faced by a private data custodian who wants to sell population statistics. Our model extends \citet[GR hereafter]{Ghosh:Auction:GEB:2015}, who consider the problem of a data custodian, or producer, with legal possession of confidential data that was originally provided by data owners. The custodian wants to sell population statistics based on the confidential data to data users, who need the statistical summaries sold by the custodian to improve decision-making. We formalize the tradeoff between privacy and accuracy by assuming the custodian publishes using a differentially private mechanism. Operating this mechanism to publish statistics with a given level of accuracy requires the data owners to incur a known and quantifiable loss of privacy. \citet{Ghosh:Auction:GEB:2015} establish a minimum-cost method for purchasing privacy-loss rights from the data owners. 
Unlike Ghosh and Roth, who treat the demand for accuracy as exogenous, we assume an endogenous demand for data accuracy, and focus on its implications for the efficiency of private provision.
The producer therefore balances the demand for statistical accuracy against a demand for privacy protection. We model consumers who have heterogeneous preferences for the accuracy of the published statistical summaries, as well as for privacy protection. This formulation nests the more intuitive case in which the users of data are distinct from the population on whom data are collected.

Our model of data publication is based on \emph{differential privacy} \citep{Dwork2006,Dworketal:2006,Dworketal:JPC:2017}, which has been adopted by tech companies like Google and Apple, as well as by the U.S. Census Bureau.\footnote{See, for example, \citet{Erlingsson2014},  \citet{Apple:Learning:2017}, and \citet{AbowdSchmutte:Privacy:AER} regarding applications of differential privacy at Google, Apple, and Census respectively.}
Differential privacy is an approach to publishing statistical summaries from confidential data sources that allows the publisher to make explicit, mathematically rigorous, statements about how much privacy---measured as a quantity---is lost with each publication. Furthermore, differentially private publications can induce an explicit positive relationship between accuracy of the published data and the amount of privacy loss.\footnote{For a non-technical introduction to differential privacy, see 
%\citet{Nissim:DPNonTech:WP:2018}. 
\citet{WoodVanderbiltJ.Entertain.Technol.Law2018}
% NOTE: The more recent Wood et al?
See \citet{Heffetz2014} for an introduction targeted toward economists. For a more comprehensive treatment, see \citet{Dwork:Roth:journal:version:2014}.}

 Crucially, we model both privacy protection and accuracy as public goods. Thus, it is not \emph{a priori} obvious whether the private provider will provide too much or too little privacy protection. Data accuracy is a public good, since any consumer may access and use the published data without reducing its accuracy for some other consumer (it is non-rival) and no consumer can block another consumer's use (it is non-excludable). 
 % Note: for non-economists, do we need a reference here?
 In plain English, all persons can learn and benefit from the use of high-quality data by others, and they can also access those data directly themselves. They value what they learn. And they understand that what they learn is more useful if it is more accurate. Privacy protection is also a public good because all individuals in the database benefit from the same level of privacy protection embodied in the producer's data publication process, an implication of the Ghosh-Roth mechanism (non-rivalry in consumption for privacy protection). 

We find that private provision results in suboptimally low data accuracy. As in Samuelson's classic model \citep{samuelson:1954}, the external benefit of data accuracy to all consumers is not captured by the willingness-to-pay of the consumer with the greatest private value. By contrast, the demand for privacy protection is derived from the data provider's cost-minimization problem. The provider buys just enough data-use rights (privacy loss) to sell the data accuracy to the consumer with the highest valuation. All other consumers use the published data for free.\footnote{Study of this case may be of special interest for some business-data collection for industries with a small number of dominant organizations.}

While the suboptimality of private provision of public goods is well-understood \citep{Spence:Monopoly:Bell:1975}, modeling the origin and nature of suboptimality in the market for population statistics is not. Given the adoption of differential privacy by the U.S. Census Bureau,
% Need reference
and the increasing demand for public data products from tech companies, it is important to consider how markets might, and might not, appropriately balance society's interests in privacy protection and data quality \citep{AbowdSchmutte:Privacy:AER}. Our paper is also broadly related to recent work in the economics of privacy \citep{acquisti:taylor:wagman:2015,Heffetz2014,Goldfarb2015}, which focuses on the role of privacy in facilitating the efficient use of customer data. Few papers have considered the economic tradeoff between privacy protection and data quality in the production of population statistics.  \citet{Ghosh:Auction:GEB:2015}, on which we build, and related papers in electronic commerce \citep{Li2014} assume the demand for accuracy is exogenous. In some settings, this is appropriate---for example, when a company is mining its customer data for internal use. Publishing data summaries is, as we now show, another matter altogether. 

%% file: preliminaries.tex
%TCIDATA{Version=5.50.0.2960}
%TCIDATA{LaTeXparent=0,0,sw-edit.tex}

% -*- latex -*-
%
% Time-stamp: <>
%              Automatically adjusted if using Xemacs
%              Please adjust manually if using other editors
%
%preliminaries.tex
% Part of abowd-schmutte-privacy.tex

This section provides our formal definitions of privacy and data accuracy. Our definitions are based on a computer science literature studying formal privacy, and so may be unfamiliar to economists. Our summary draws on several sources to which we refer the reader who is interested in more details %
\citep{Hardt:Multiplicative:FOCS10,Dwork:Roth:journal:version:2014,Wasserman:Statistical:JASA:2010,Heffetz2014,AbowdSchmutte:Privacy:AER}. Our notation follows \citet{Dwork:Roth:journal:version:2014}.

We introduce the notion of differential privacy, which is key to understanding our analysis. Differentially private data publications do not allow an outsider to learn ``too much'' about any individual data record based on statistical summaries of the full database.
For our purposes, this framework is useful because differential privacy tells us, for any level of data accuracy, how much privacy loss an efficient provider must be willing to tolerate.

\subsection{Databases and Queries}
A data custodian (e.g.,\ Facebook, Google, the U.S.\ Census Bureau) possesses a database, $D$. 
Think of $D$ as a table in which each row represents information for a single individual and each column
represents a single characteristic to be measured. The database $D$ contains
$N$ rows. 
% The set $\chi$ describes all possible values the variables in the
% columns of the database can take. Therefore any row that appears in the
% database is an element of $\chi$.\footnote{%
% In statistics, $\chi$ is the sample space. All structural zeros
% (combinations of values that are deemed impossible a priori) are removed
% from $\chi$. For example, if the variables recorded in the database are a
% binary indicator for gender, $g\in \{0,1\}$, and a categorical indicator for six
% different levels of program eligibility, $s\in \{1,\ldots ,6\}$, then $\chi
% =\{0,1\}\times \{1,\ldots ,6\}$, and $|\chi|=12$.
% If the pair (0,6) is impossible, then $\chi=\{0,1\}\times \{1,\ldots ,5\} \cup {(1,6)}$, and $|\chi|=11$.
% } 
We assume all variables are discrete and finite-valued, but this is not restrictive since continuous data are
always given discrete, finite representations when recorded on search logs, email contents, network features, social media posts, censuses, surveys, or administrative record systems.

The notion of a \emph{neighboring database} is crucial for the definition of differential privacy. Differential privacy captures the idea that published output should not change ``too much'' based on a single data item. We say database $D^{\prime}$ is a neighboring database of $D$ if $D^{\prime}$ can be obtained by modifying a single row in $D$. We denote by $\mathcal{D}$ the set of all admissible databases.\footnote{Formally, $D$ and $D^{\prime}$ are neighbors if the $\ell_1$-norm of the difference in their histogram representations is 2. See Appendix \ref{app:def_neighbors}.} %COMMENT JMA: This is new. Appendix A.1 does not define neighbors for bounded-DP, it is the definition for unbounded-DP. I think if you modify a single element of a single row, changing to a random legal value in the sample space, that the bounded-DP $l_1$ norm in appendix A.1. changes by 2, not 1, because two elements of the histogram must change.

Data users are interested in learning answers to a \emph{database query}. A query is a function, $Q:\mathcal{D}\rightarrow \mathbb{R}^{K}$ that maps input databases, $D\in\mathcal{D}$ to a vector in $\mathbb{R}^{K}$. The concept of a database query can admit standard population statistics, like subgroup counts, means, variances, and so on, but is much broader. The case we consider in this paper focuses on publication of a single-valued query, but note that in general the query answer $Q(D)$ may be vector-valued.

% \subsubsection{Queries}
% A \textit{linear query} is a mapping $f:[0,1]^{|\chi |}\times \mathbb{Z}%
% ^{\ast |\chi |}\rightarrow \mathbb{R}^{\ast }$ such that $f(m,x)=m^{T}x$
% where $x\in \mathbb{Z}^{\ast |\chi |}$, $m\in \lbrack 0,1]^{|\chi |}$, and $\mathbb{R}^{\ast }$ is the set of
% non-negative real numbers. A
% \textit{counting query} is a special case in which $m_{i}$ is restricted to
% take a value in $\{0,1\}$. Counting queries return the number of
% observations that satisfy particular conditions. They are the tool an
% analyst would use to calculate multidimensional margins from the fully-saturated contingency
% table representation of the database, which is identical to our histogram
% representation. A \textit{normalized linear query} is a mapping $%
% f:[0,1]^{|\chi |}\times \mathbb{Z}^{\ast |\chi |}\rightarrow \lbrack
% 0,1\rbrack$ such that if $\tilde{f}$ is a linear query, then $f(m,x)=\tilde{f}%
% (m,x)/||x||_{1}$.

% We model queries about population proportions rather than
% counts. These correspond to the proportions from a saturated contingency table.
% To that end, we work with
% normalized linear queries unless otherwise specified. The use of
% normalization is not restrictive. It affects the functional form of
% privacy and accuracy bounds only through their dependence on the database size
% $||x||_{1}$. Any bound stated in terms of the unnormalized histograms and
% queries can be restated in terms of normalized histograms and queries.

\subsection{Query Release Mechanisms, Privacy, and Accuracy}
%We model the data release mechanism as a randomized algorithm.
% Note: do we need a footnote or reference for economists to define this terms?
The data
curator operates a query release mechanism that provides answers to queries $Q$ given a database $D$.

\begin{definition}[Query Release Mechanism]
\label{def:query_mechanism} Let $\mathcal{Q}$ be a set of admissible single-valued queries.
% and $|\mathcal{Q}|$ the number of queries to be answered. 
A query release mechanism $M$ is a random function $M:\mathcal{D}\times \mathcal{Q}\rightarrow \mathbb{R}$ whose inputs are a database $D$ and a query $Q$.
The mechanism output is a probabilistic response to the query.
The probability of observing $B\subseteq \mathbb{R}$
is $\Pr \left[ M(D,Q)\in B|D,Q\right] $,
the conditional probability, given $D$ and $Q$,
that the published query answer is in $B\in \mathcal{B}$%
, where $\mathcal{B}$ are the measurable subsets of $\mathbb{R}$.
\end{definition}

%COMMENT JMA: This is false generality. We never consider the vector-valued query case, and we rely on proofs below that are explicitly single-value queries. I think we should say that $Q$ is a single element of $\mathcal{Q}$ and $B\subseteq \mathbb{R}$.

\subsubsection*{Differential Privacy}

Our definitions of differential privacy and accuracy for the query release
mechanism follow \citet{Dworketal:2006} and \citet{Kifer:2011:NFL:1989323.1989345}. %\citet{Hardt:Multiplicative:FOCS10} and %
%\citet{Dwork:Roth:journal:version:2014}.
\footnote{Following the setup in \citet{Ghosh:Auction:GEB:2015}, we are using the variant of differential privacy now known as \emph{bounded} differential privacy. This means that the total number of records in the confidential database, called $N$ below, is publicly known.} %COMMENT: or we can switch to bounded DP, or switch to unnormalized queries, and ask total and bit=1.

\begin{definition}[$\protect\varepsilon$-differential privacy]
\label{def:dif_priv} Query release mechanism $M$ satisfies $\varepsilon$%
-differential privacy if for $\varepsilon > 0$, for all pairs of neighboring databases $D,D^{\prime }$, all queries $Q\in\mathcal{Q}$, and all $B\in \mathcal{B}$
\begin{equation*}
\Pr \left[ M(D,Q)\in B |D,Q\right] \leq e^{\varepsilon }\Pr \left[
M(D^{\prime },Q)\in B |D^{\prime},Q\right],
\end{equation*}%
where $\mathcal{B}$ are the measurable subsets of $\mathbb{R}$, and the randomness in $M$ is due exclusively to the mechanism and not the process generating the database $D$.
% , $\lbrack 0,1\rbrack^{|\mathcal{Q}|}$.
\end{definition}

\subsubsection*{Accuracy}
We next define our measure of accuracy. For any query, $Q\in\mathcal{Q}$, the query release mechanism returns an answer, $a$, that depends on the input database, the content of the query response, and the randomization induced by the query release mechanism.
\begin{definition}[$(\protect\alpha ,\protect\beta )$-accuracy]
\label{def:acc} Query release mechanism $M$ satisfies $(\alpha ,\beta )$%
-accuracy if for $Q\in\mathcal{Q}$ and $a$ output from $M(D,Q)$,
%\begin{equation*}
%\min_{1\le k\le K}\left\{ \Pr %\left[ \vert a_{k} - Q(D)_{k} \vert %\leq \alpha |D,Q %
%\right] \right\} \geq 1-\beta.
%\end{equation*}
%$$ \min_{1 \le k \le K} \; \mbox{Pr}\Big( \, |a_k-Q(D)_k| \le \alpha \; \Big| \; D,Q\Big) \; \ge \; 1-\beta $$
$$\mbox{Pr}\Big( \, |a-Q(D)| \le \alpha \; \Big| \; D,Q\Big) \; \ge \; 1-\beta $$
where $a, Q(D) \in \mathbb{R}$.
%where $k$ indexes the components of the mechanism output vector in $\mathbb{R}^{K}$, and $a_k$ and $Q(D)_{k}$ are the corresponding elements of the published and confidential outputs.  
\end{definition}
This definition guarantees that the error in the answer provided by the mechanism is bounded above by $\alpha $ with probability $(1-\beta )$.\footnote{This definition also appears in a more general form in \citet{Gupta:2012:ICP:2238936.2238961}.} The probabilities in the definition of $\left( \alpha ,\beta \right) $-accuracy are induced by the query release mechanism.

\subsection{Example}\label{runex}
%We offer two examples to characterize the problem stylized by the model.
To illustrate the problem stylized by the model, consider the following scenario.

%\subsubsection*{Example 1: Publication of ethnicity} 
% The US Census Bureau collects and publishes summaries of the ethnicity of respondents to the Decennial Census. For example, 
Under Public Law 94-171, the U.S.~Census Bureau publishes the number of individuals of Hispanic origin in each census block. Many blocks have small populations. Publishing the size of the Hispanic population without statistical disclosure limitation can lead to disclosure of the ethnicity of individuals in that block.

Framed in terms of the model, the database of interest, $D$, has one row for each person in a given block, and includes a binary indicator of their Hispanic origin. A neighboring database, $D^{\prime}$, has the same rows, but the Hispanic origin is changed for exactly one entry. The query of interest, $Q(D)$, is the proportion of individuals of Hispanic origin in the block. If the Census Bureau publishes the answer to $Q(D)$ exactly, then an attacker who knows the Hispanic origin of all but one individual can learn the origin of the remaining individual with certainty. If, instead, it publishes a noisy proportion under a differentially private query release mechanism, $M(D,Q)$, such an attacker will remain uncertain of the origin of the remaining individual. The question is whether the noisy proportion still measures the true proportion with sufficient accuracy. 

% Note: this seems short.
%\subsubsection*{Example 2: Publication of the unemployment rate} 
In the preceding example, the attribute of interest is a binary indicator. This may seem to be a restrictive assumption. As we noted earlier, continuous data are
generally given discrete, finite representations when recorded in databases. For instance, Public Law 94-171 also requires publication of the number of individuals in blocks and tracts, classified by age, a continuous feature. However, in the database $D$, age is discretized to single years, and for block-level tabulations, to year ranges. An 52-year old individual is thus recorded as the binary response to the query ``age = 52''. 
%
% Note: does not seem intuitive
%
%and through the rest of the paper, we suppose the database of interest has data on a single binary attribute. For example, a binary indicator of an individual's unemployment status published from the Current Population Survey (CPS) is derived from a recode of a set of individual characteristics, including age, employment status, and reported job search behavior. 
More generally, our model applies for publication of any predicate query that asks whether an individual's characteristics satisfy a set of binary conditions.

% Several laws govern the Bureau of Labor Statistics handling of employment data including the Confidential Information Protection and Statistical Efficiency act, the Workforce Investment Act, and the Privacy Act.

% In terms of our model, the data curator holds a database $V$ in which each individual's record consists of several attributes. The analysis is based on an alternative database $D$ that has one row for each individual, and a binary unemployment indicator derived from the attributes in $V$. The query of interest, $Q(D)$, is the unemployment rate. As above, if the curator publishes a noisy unemployment rate using a differentially private release mechanism, an attacker is prevented from learning ``too much'' about the unemployment status of any individual.

%\noindent{\emph{Example 3: Publication of }}

%%% Local Variables:
%%% mode: latex
%%% TeX-master: abowd-schmutte-privacy
%%% End:

%% file: suboptimality.tex
%TCIDATA{Version=5.50.0.2960}{def:acc}
%TCIDATA{LaTeXparent=0,0,sw-edit.tex}
% -*- latex -*-
%
% suboptimality.tex
% Part of abowd-schmutte-privacy.tex
In this section, we model a data provider selling public statistics produced according to a differentially private mechanism by purchasing rights to use records in an underlying confidential database. Since accuracy and privacy protection are both public goods, the consequences of private provision are theoretically ambiguous until further structure is placed on the model. Given the structure below, we prove that too little data accuracy and too much privacy will be supplied by a private provider compared to the social welfare maximizing solution.

\subsection{Model Setup}
Each of $N$ private individuals possesses a single bit of information, $b_{i}$, that is already
stored in a database maintained by a trusted curator.\footnote{%
Trusted curator can have a variety of meanings. We mean that the database is
held by an entity, governmental or private, whose legal authority to hold
the data is not challenged and whose physical data security is adequate to
prevent privacy breaches due to theft of the confidential data themselves.
We do not model how the trusted curator got possession of the data, but we
do restrict all publications based on these data to use statistics produced
by a query release mechanism that meets the same privacy and confidentiality
constraints.}
In addition to their private information, each individual is endowed with
income, $y_{i}$.

Individuals each consume one unit of the published statistic, which has
accuracy $I$ defined in terms of $(\alpha ,\beta )$-accuracy, that is $%
I=(1-\alpha )$. Since $I$ is a public good, all consumers enjoy the benefits
of $I$, but each consumer is charged the market price $p_{I}$, to be
determined within the model, for her \textquotedblleft
share\textquotedblright\ of $I$, which we denote $I_{i}$, and the balance of
the public good, which we denote $I^{\symbol{126}i}$ is paid for by the
other consumers. Thus, $I=I_{i}+I^{\symbol{126}i}$ for all consumers.

The preferences of consumer $i$ are given by the indirect utility function
\begin{equation}
v_{i}\left( y_{i},\varepsilon _{i},I_{i},I^{\symbol{126}i}\right) =\ln
y_{i}+p_{\varepsilon }\varepsilon _{i}-\gamma _{i}\varepsilon _{i}+\eta
_{i}\left( I_{i}+I^{\symbol{126}i}\right) -p_{I}I_{i}.  \label{eqn:linear2}
\end{equation}%
Equation~(\ref{eqn:linear2}) implies that preferences are quasilinear in
data accuracy, $I$, privacy loss, $\varepsilon _{i}$, and log income, $\ln
y_{i}$.\footnote{%
In this section, we keep the description of preferences for data accuracy
and privacy protection as close as possible to the 
%original Ghosh and Roth
GR
specification. They allow for the possibility that algorithms exist that can
provide differential privacy protection that varies with $i$; hence $%
\varepsilon _{i}$ appears in equation~(\ref{eqn:linear2}). They subsequently
prove that $\varepsilon _{i}=\varepsilon $ for all $i$ in their Theorem 3.3.}
We incorporate income and accuracy in the utility function because
they are required for the arguments in this section. 

The term $%
p_{\varepsilon }$ is the common price per unit of privacy, also to be
determined by the model. The receipt $p_{\varepsilon }\varepsilon _{i}$
represents the total payment an individual receives if her bit is used in an
$\varepsilon $-differentially private mechanism. The individual's marginal
preferences for data accuracy (a \textquotedblleft good\textquotedblright )
and privacy loss (a \textquotedblleft bad,\textquotedblright\ really an
input here), $\left( \eta_{i},\gamma _{i}\right) >0,$ are not known to the
data provider, but their population distributions are public information.
Therefore, the mechanism for procuring privacy has to be individually
rational and dominant-strategy truthful.

We do not include any explicit interaction between the publication of statistical data and the market for private goods. This assumption is not without consequence, and we make it to facilitate exposition of our key point: that data accuracy may be under-provided due to its public-good properties. Violations of privacy might affect the goods market through targeted advertising and price discrimination. The accuracy of public statistics may also spill over to the goods market by making firms more efficient. These are topics for future work.

\subsection{The Cost of Producing Data Accuracy}
A supplier of statistical information wants to produce an $\left( \alpha
,\beta \right) $-accurate estimate, $\hat{s}$, of the population statistic%
\begin{equation}
    s=\frac{1}{N}\sum_{i=1}^{N}b_{i}  \label{eqn:s_def}
\end{equation}%
\textit{i.e.}, a normalized query estimating the proportion of individuals
with the property encoded in $b_{i}$. This property could be something highly sensitive, such as the individual's citizenship status, sexual orientation, or whether she suffers from a particular health condition.

Theorems 3.1 and 3.3 in \citet{Ghosh:Auction:GEB:2015} prove that publishing
the statistic
\begin{equation}
\hat{s}=\frac{1}{N}\left[ \sum_{i=1}^{H}b_{i}+\frac{\alpha N}{2\left(1/2+\ln \frac{1}{\beta} \right)} +%
\func{Lap}\left( \frac{1}{\varepsilon }\right) \right] \label{eqn:GRquery}
\end{equation}%
provides $\left( \alpha ,\beta\right) $-accuracy, and requires a privacy loss of $%
\varepsilon _{i}=\varepsilon =\frac{1/2+\ln{\left(1/\beta\right)}}{\alpha N}$ from $H=N-\frac{%
\alpha N}{1/2+\ln{\left(1/\beta\right)}}$ members of the population.  $\func{Lap}\left(\frac{1}{\varepsilon }\right) $
represents a draw from the Laplace distribution with mean $0$ and scale
parameter $\frac{1}{\varepsilon }$.

Purchasing the data-use rights from the $H$ least
privacy-loving members of the population; \textit{i.e.}, those with the
smallest $\gamma _{i}$, is the minimum-cost, envy-free implementation
mechanism \citep{Ghosh:Auction:GEB:2015}.\footnote{%
We note for completeness that the statistic $\hat{s}$, while computed on only $H$ cases from 
the population of $N$, is evaluated relative to the population quantity $s$. GR use the same accuracy measure 
as we do; namely Definition \ref{def:acc} with a single query in the query set, although they assume $\beta=\frac{1}{3}$ throughout. We restrict the choice of $\beta$ to $\beta < 1/\left(1+\sqrt{e}\right)$; the threshold required by the proof technique in GR, Theorem 3.1.
Statisticians often use mean squared error instead of the absolute error embodied in this definition. 
Nevertheless, the statistic $\hat{s}$ also trades-off bias and variance relative
to the correct population statistic. The term $\alpha N/\left[2\left(1/2 + \ln{\left(1/\beta\right)}\right)\right]$ is a bias correction.}
%\citet{Ghosh:Auction:GEB:2015}
GR provide two mechanisms for implementing their VCG auction.
We rely on their mechanism \textit{MinCostAuction} and the properties given in
their Proposition 4.5. See Appendix \ref{app:GR} for additional details.\footnote{Note that the result in equation~(\ref{eqn:GRquery}) holds regardless of any correlation between privacy preferences and data values. That is, even if it is biased, the summary produced using data from those consumers with the lowest privacy preferences still satisfies the accuracy guarantee.}

We now derive the producer's problem of providing the statistic for a given
level of data accuracy, $I$. If $%
p_{\varepsilon }$ is the payment per unit of privacy loss, the total cost of
production is $c(I)=p_{\varepsilon }H\varepsilon $, where the right-hand
side terms can be defined in terms of $I$ as follows. Using the arguments
above, the producer must purchase from $H(I)$ consumers the right to use
their data to compute $\hat{s}$. Then,
\begin{equation}
    H(I)=N-\frac{(1-I)N}{1/2+\ln{\left(1/\beta\right)}}.  \label{eqn:H_def}
\end{equation}%
Under the VCG mechanism, the price of privacy loss must be $p_{\varepsilon
}=Q\left( \frac{H(I)}{N}\right) $, where $Q$ is the quantile function with
respect to the population distribution of privacy preferences, $F_{\gamma }$%
. The lowest price at which the fraction $\frac{H(I)}{N}$ of consumers do
better by selling the right to use their bit, $b_{i}$, with $\varepsilon
\left( I\right) $ units of differential privacy is $p_{\varepsilon }$. $H(I)$
is increasing in $I$. The total cost of producing $I$ is
\begin{equation}
    C^{VCG}(I)=Q\left( \frac{H(I)}{N}\right) H(I)\varepsilon (I),
    \label{eqn:c_def}
\end{equation}%
where the production technology derived by GR implies%
\begin{equation}
    \varepsilon (I)=\frac{1/2+\ln{\left(1/\beta\right)}}{(1-I)N}.  \label{eqn:e_def}
\end{equation}

\subsection{Example}

% The strategy of acquiring data-use rights from the least privacy concerned individuals requires high participation to achieve even modest accuracy guarantees. This is not too surprising. 
The results of \citet{Ghosh:Auction:GEB:2015} hold even with an arbitrary correlation between privacy preferences and measured characteristics. Obtaining accuracy in the presence of potentially extreme selection bias makes it necessary to count almost everyone. In practice, the costs of publication are determined by the level of privacy-loss required, and the preferences of someone with extreme aversion to privacy loss. 

Recall Example 1 from Section \ref{runex}, which involved computing the share of the Hispanic population in a census block. If an analyst requires $(0.2, 0.1)$-accuracy in the estimate, then she must purchase privacy loss of $\epsilon \approx 14/N$ from $H=0.93N$ people, or $93$ percent of the block population. Note that the required privacy loss, $\epsilon$, vanishes as the block size $N$ increases. To obtain a stronger guarantee of $(0.05,0.05)$-accuracy would require purchasing $\epsilon \approx 70/N$ from $H=0.99N$ people. A very weak guarantee of $(0.4, 1/3)$-accuracy only requires privacy loss of $\epsilon \approx 4/N$ from $H=0.75N$ people.

\section{Suboptimality of Private Provision\label%
{subsec:competitive}}

% \subsection{Private, Competitive Supply of Data Accuracy}
Suppose a private profit-maximizing, price-taking, firm sells $\hat{s}$ with
accuracy $\left( \alpha ,\beta\right) $, that is, with data accuracy $%
I$ at price $p_{I}$. Then, profits $P\left( I\right) $ are%
\begin{equation*}
    P\left( I\right) =p_{I}I-C^{VCG}(I).
\end{equation*}%
If it sells at all, it will produce $I$ to satisfy the first-order condition
$P^{\prime }\left( I^{VCG}\right) =0$ implying%
\begin{equation}
    p_{I}=Q\left( \frac{H(I)}{N}\right) H(I)\varepsilon ^{\prime }(I)+\left[
    Q\left( \frac{H(I)}{N}\right) +Q^{\prime }\left( \frac{H(I)}{N}\right)
    \left( \frac{H(I)}{N}\right) \right] H^{\prime }(I)\varepsilon (I)
    \label{eq:p_private}
\end{equation}%
where the solution is evaluated at $I^{VCG}$.\footnote{%
The second order condition is $P^{\prime \prime }\left( I^{VCG}\right) <0$,
or $\frac{d^{2}C^{VCG}\left( I\right) }{dI^{2}}>0$. The only term in the
second derivative of $C^{VCG}\left( I\right) $ that is not unambiguously
positive is $\frac{H\left( I\right) H^{\prime }\left( I\right)
^{2}\varepsilon \left( I\right) }{N^{2}}Q^{\prime \prime }\left( \frac{%
H\left( I\right) }{N}\right) $. We assume that this term is dominated by the
other, always positive, terms in the second derivative. Sufficient
conditions are that $Q\left( {}\right) $ is the quantile function from the
log-normal distribution or the
quantile function from a finite mixture of normals, and that $\frac{H\left(
I\right) }{N}$ is sufficiently large; \textit{e.g.}, large enough so that if
$Q\left( {}\right) $ is the quantile function from the $\ln N\left( \mu
,\sigma ^{2}\right) $ distribution, $Q^{\ast \prime \prime }\left( \frac{%
H\left( I\right) }{N}\right) +\sigma ^{2}Q^{\ast \prime }\left( \frac{%
H\left( I\right) }{N}\right) ^{2}\geq 0$, where $Q^{\ast }\left( {}\right) $
is the standard normal quantile function.} The price of data accuracy is
equal to the marginal cost of increasing the amount of privacy
protection--data-use rights--that must be purchased. There are two terms.
The first term is the increment to marginal cost from increasing the amount
each privacy-right seller must be paid because $\varepsilon $ has been
marginally increased, thus reducing privacy protection for all. The second
term is the increment to marginal cost from increasing the number of people
from whom data-use rights with privacy protection $\varepsilon $ must be
purchased. As long as the cost function is strictly increasing and convex,
the existence and uniqueness of a solution is guaranteed.

\subsection{Competitive Market Equilibrium}
At market price $p_{I}$, consumer $i$'s willingness to pay for data accuracy
will be given by solving%
\begin{equation}
    \max_{I_{i}\geq 0}\eta _{i}\left( I^{\symbol{126}i}+I_{i}\right) -p_{I}I_{i}
    \label{eqn:umax}
\end{equation}%
where $I^{\symbol{126}i}$ is the amount of data accuracy provided from the
payments by all other consumers, as noted above. Consumer $i$'s willingness
to pay is non-negative if, and only if, $\eta _{i}\geq p_{I}$; that is, if
the marginal utility from increasing $I$ exceeds the price. If there exists
at least one consumer for whom $\eta _{i}\geq p_{I}$, then the solution to
equation (\ref{eq:p_private}) is attained for $I^{VCG}>0.$

We next show that there is only one such consumer. It is straightforward to
verify that the consumers are playing a classic free-rider game \citep[pp.
361-363]{mas1995microeconomic}. In the competitive equilibrium, the only
person willing to pay for the public good is one with the maximum value of $%
\eta _{i}$. All others will purchase zero data accuracy but still consume
the data accuracy purchased by this lone consumer. Specifically, the
equilibrium price and data accuracy will satisfy%
\begin{equation*}
    p_{I}=\bar{\eta}=\frac{dC^{VCG}\left( I^{VCG}\right) }{dI},
\end{equation*}%
where $\bar{\eta}$ is the maximum value of $\eta _{i}$ in the
population--the taste for accuracy of the person who desires it the most.
However, the Pareto optimal consumption of data accuracy, $I^{0},$ solves
\begin{equation}
    \sum_{i=1}^{N}\eta _{i}=\frac{dC^{VCG}\left( I^{0}\right) }{dI}.
    \label{eqn:pareto_optimality}
\end{equation}%
Marginal cost is positive, $\frac{dC^{VCG}\left( I^{0}\right) }{dI}>0$, and $%
\sum_{i=1}^{N}\eta _{i} > \bar{\eta}$; therefore, data accuracy will be
under-provided by a competitive supplier when data accuracy is a public good
as long as marginal cost is increasing, which we prove below. More
succinctly, $I^{VCG} < I^{0}$. Therefore, privacy protection must be
over-provided, $\varepsilon ^{VCG} < \varepsilon ^{0}$, by equation (\ref%
{eqn:e_def}).\footnote{%
The reader is reminded that a smaller $\varepsilon $ implies more privacy
protection. It is also worth commenting that in the GR formulation the
single consumer with positive willingness to pay is the entity running the
VCG auction. That person is buying data-use rights from the other consumers,
computing the statistic for publication, then releasing the statistic so
that all other consumers may use it. That is why we have modeled this as a
public good. Our formulation is fully consistent with GR's scientist seeking data for a
grant-supported publication.}

\subsection{The Price-discriminating Monopsonist Provider of Data accuracy}
Now consider the problem of a single private data provider who produces $%
\hat{s}$ with accuracy $\left( \alpha ,\beta\right) $ using the same
technology as in equations (\ref{eqn:c_def}) and (\ref{eqn:e_def}). We now
allow the producer to price-discriminate in the acquisition of data-use
rights--that is, the private data-accuracy supplier is a price discriminating
monopsonist. This relaxes the assumptions
of the VCG mechanism in %\citet{Ghosh:Auction:GEB:2015}
GR to allow for the unrealistic possibility that the
data accuracy provider knows the population values of $\gamma _{i}$. They
acknowledge this theoretical possibility when discussing the individual
rationality and dominant-strategy truthful requirements of their mechanism.
They reject it as unrealistic, and we agree. We are considering this
possibility to show that even when the private data-accuracy provider is
allowed to acquire data-use rights with a lower cost strategy than the VCG
mechanism, data accuracy will still be under-provided.

The producer must decide how many data-use rights (and associated privacy
loss $\varepsilon $, the same value for all $i$) to purchase from each
member of the database, or, equivalently, how much to charge members of the
database to opt out of participation in the mechanism for computing the
statistic. (They cannot opt out of the database.) Let $\pi \in \left\{
0,1\right\} ^{N}$ be the participation vector. Using the Lindahl approach,
let $p_{\varepsilon _{i}}^{L}$ be the price that satisfies, for each
consumer $i$,%
\begin{equation}
p_{\varepsilon _{i}}^{L}\leq \gamma _{i},\text{with equality if }\pi _{i}=1.
\label{eqn:lindahl}
\end{equation}%
Equation~(\ref{eqn:lindahl}) says that the Lindahl prices are those such
that the choice of $\varepsilon $ is exactly the value that each individual
would optimally choose on her own. Even with our assumption of linear
preferences, the Lindahl prices are unique for every consumer who
participates in the mechanism for computing the statistic.

Given a target data accuracy of $I=(1-\alpha )$, the producer's cost
minimization problem is the linear program%
\begin{equation}
C^{L}\left( I\right) =\min_{\pi }\left( \sum_{i=1}^{N}\pi _{i}p_{\varepsilon
_{i}}^{L}\right) \varepsilon  \label{eqn:lindahl_cost}
\end{equation}%
subject to%
\begin{equation*}
\sum_{i=1}^{N}\pi _{i}=N-\frac{(1-I)N}{1/2+\ln{\left(1/\beta\right)}}\text{ and }\varepsilon =%
\frac{1/2+\ln{\left(1/\beta\right)}}{(1-I)N}.
\end{equation*}%
The solution is for the producer to set $\pi _{i}=1$ for the $H$ members of
the database with the smallest $p_{\varepsilon _{i}}^{L}$ and $\pi _{i}=0$,
otherwise. Note that if
\begin{equation*}
\frac{dC^{L}\left( I\right) }{dI} < \frac{dC^{VCG}(I)}{dI}
\end{equation*}%
for all $I$, which will be proven in Theorem \ref{theorem:suboptimality},
then the Lindahl purchaser of data-use rights will produce more data accuracy
at any given price of data accuracy than the VCG purchaser.

By construction, the Lindahl solution satisfies the Pareto optimality
criterion for data-use rights acquisition that%
\begin{equation}
\sum_{i=1}^{N}\pi _{i}p_{\varepsilon _{i}}^{L}=\sum_{i=1}^{N}\pi _{i}\gamma
_{i}.  \label{eqn:lindahl_PO}
\end{equation}%
Once again, the supplier implements the query response mechanism of equation~(\ref{eqn:GRquery}) with $ \frac{1/2+\ln{\left(1/\beta\right)}}{(1-I)N} $-differential privacy and $(1-I,\beta)$-accuracy but pays
each consumer differently for her data-use right. Notice that equation~(\ref{eqn:lindahl_PO}) describes the Pareto optimal privacy
% Note: avoid "whether or not" formulations...
loss whether or not one acknowledges that the privacy protection afforded by
$\varepsilon $ is non-rival, only partially excludable, and, therefore, also
a public good.

To implement the Lindahl solution, the data producer must be able to exclude
the bits, $b_{i}$, of specific individuals when computing the statistic, and
must have perfect knowledge of the every marginal disutility $\gamma _{i}$
of increasing $\varepsilon $. When this information is not available, the
producer can, and will, implement the first-best allocation by choosing a
price through the VCG auction mechanism used by GR.

% Note: why this paragraph, here?
For readers familiar with the data privacy literature, we note that the
statement that technology is given by equations (\ref{eqn:c_def}) and (\ref%
{eqn:e_def}) means that the data custodian allows the producer to purchase
data-use rights with accompanying privacy loss of $\varepsilon =\frac{%
1/2+\ln{\left(1/\beta\right)}}{(1-I)N}$ from $H\left( I\right) $ individuals for the sole
purpose of computing $\hat{s}$ via the query response mechanism in equation $%
\left( \ref{eqn:GRquery}\right) $ that is $\frac{1/2+\ln{\left(1/\beta\right)}}{(1-I)N}$%
-differentially private and achieves $(1-I,\beta)$-accuracy, which is
exactly what Ghosh and Roth prove.

\subsection{Proof of Suboptimality}

% Note: should this be broken into two theorems, one for I^VCG < I^L and one for I^VCG < I^0?

\begin{theorem}
\label{theorem:suboptimality}If preferences are given by equation (\ref%
{eqn:linear2}), the query response mechanism satisfies equation (\ref%
{eqn:e_def}) for $\varepsilon$-differential privacy with $\left( 1-I,\beta\right) $-accuracy, cost functions satisfy (\ref{eqn:c_def}) for the VCG
mechanism, and (\ref{eqn:lindahl_cost}) for the
Lindahl mechanism,
the population distribution of $\gamma $ is given by $F_{\gamma }$ (bounded,
absolutely continuous, everywhere differentiable, and with quantile function
$Q$ satisfying the conditions noted in Section \ref{subsec:competitive}),
the population distribution of $\eta $ has bounded support on $\left[ 0,\bar{%
\eta}\right] $, and the population in the database is represented as a
continuum with measure function $H$ (absolutely continuous, everywhere
differentiable, and with total measure $N$) then
$I^{VCG} < I^{L}$ and $I^{VCG} < I^{0}$, where $%
I^{0}$ is the Pareto optimal level of $I $ solving equation $\left( \ref%
{eqn:pareto_optimality}\right) $,
$I^{L}$ is the privately-provided level when using the Lindahl mechanism to procure
data-use rights
and $I^{VCG}$ is the privately-provided level when using the VCG\
procurement mechanism.

\begin{proof}
By construction, $F_{\gamma }(\gamma )$ is the distribution of Lindahl prices.
Given a target error bound $\alpha $, corresponding to data accuracy level $I=(1-\alpha )$, 
the private producer must procure data-use rights from the respondents in the confidential data 
with $\varepsilon (I)$ units of privacy protection from a measure of $H(I)$ individuals.
Define%
\begin{equation*}
p_{\varepsilon }^{\ell}=Q\left( \frac{H(I)}{N}\right),
\end{equation*}
for $\ell = VCG, L$.
Note that $p_{\varepsilon }^{\ell}$ is the disutility of privacy loss for the marginal participant in the VCG\ and Lindahl mechanisms, respectively.
The total cost of producing $I=(1-\alpha )$ using the VCG\ mechanism is equation (\ref{eqn:c_def}):
\begin{equation*}
C^{VCG}(I)=Q\left( \frac{H(I)}{N}\right) H(I)\varepsilon (I).
\end{equation*}%
while the total cost of implementing the Lindahl mechanism is equation (\ref%
{eqn:lindahl_cost}):%
\begin{equation*}
C^{L}(I)=\left( N\int_{0}^{Q\left( \frac{H(I)}{N}\right) }\gamma dF_{\gamma
}(\gamma )\right) \varepsilon (I).
\end{equation*}%
Using integration by parts and the properties of the quantile function,%
\begin{eqnarray*}
C^{L}(I) &=&\left[ Q\left( \frac{H(I)}{N}\right) F_{\gamma }\left( Q\left(
\frac{H(I)}{N}\right) \right) -\int_{0}^{Q\left( \frac{H(I)}{N}\right)
}F_{\gamma }(\gamma )d\gamma \right] N\varepsilon (I) \\
&=&\left[ Q\left( \frac{H(I)}{N}\right) H(I)-N\int_{0}^{Q\left( \frac{H(I)}{N%
}\right) }F_{\gamma }(\gamma )d\gamma \right] \varepsilon (I).
\end{eqnarray*}%
Differentiating with respect to $I$,%
\begin{equation*}
\frac{dC^{L}(I)}{dI}=\left[ Q\left( \frac{H(I)}{N}\right)
H(I)-N\int_{0}^{Q\left( \frac{H(I)}{N}\right) }F_{\gamma }(\gamma )d\gamma %
\right] \varepsilon ^{\prime }(I)+Q\left( \frac{H(I)}{N}\right) H^{\prime
}(I)\varepsilon (I).
 \end{equation*}%
The corresponding expression for $C^{VCG}(I)$ is %
\begin{equation*}
\frac{dC^{VCG}(I)}{dI}=Q\left( \frac{H(I)}{N}\right) H(I)\varepsilon
^{\prime }(I)+\left[ Q\left( \frac{H(I)}{N}\right) +Q^{\prime }\left( \frac{%
H(I)}{N}\right) \frac{H(I)}{N}\right] H^{\prime }(I)\varepsilon (I).
\end{equation*}%
Comparison of the preceding marginal cost expressions establishes that 
$0< \frac{dC^{L}(I)}{dI} < \frac{dC^{VCG}(I)}{dI}$ for all $I$, since
$N\int_{0}^{Q\left(\frac{H(I)}{N}\right)}F_{\gamma}(\gamma)d\gamma > 0$,
$\varepsilon^{\prime}(I)>0$, $H^{\prime }(I)>0$, and $Q^{\prime }\left( {}\right) >0$.
The results stated in the theorem follow by using the equilibrium price for the
private market sale of $I$, which is $p_{I}$ in equation (\ref{eqn:linear2}),%
\begin{equation*}
	p_{I}=\bar{\eta}=\frac{dC^{L}(I^{L})}{dI}=\frac{dC^{VCG}(I^{VCG})}{dI}.
\end{equation*}%
Hence, $I^{VCG} < I^{L}$, since $\frac{dC^{L}(I^L)}{dI} < \frac{dC^{VCG}(I^{L})}{dI}$ and the conditions on $Q$ imply that $\frac{d^{2}C^{VCG}(I)}{%
dI^{2}}> 0$.
Likewise, $I^{VCG} < I^{0}$, since
$\sum_{i=1}^{N}\eta _{i} > \bar{\eta}$, and $\frac{d^{2}C^{VCG}(I)}{%
dI^{2}}> 0$.
\end{proof}

\end{theorem}

%%% Local Variables:
%%% mode: latex
%%% TeX-master: abowd-schmutte-privacy
%%% End:

%% file: conclusion.tex
The concept of differential privacy allows a natural
interpretation of privacy protection as a commodity over which individuals
might have preferences. In many important contexts, privacy protection and
data accuracy are not purely private commodities. When both are public goods, the
market allocations might not be optimal. The solution to the social planning problem that optimally provides both
public goods--data accuracy and privacy protection--delivers more data
accuracy, but less privacy protection, than the VCG mechanism for private-provision of data. The reason is that the VCG mechanism for procuring data-use rights ignores the public-good nature of the statistics that are published after a citizen sells the right to use her private data in those publications. 

This matters because the demand for public data is greater than ever, while funding for statistical agencies has been relatively stagnant. It is increasingly likely that data users will turn to private companies to obtain the information they demand. Our results suggest that, while the policy debate has centered on regulating the privacy loss from this trend, it is also important to counterbalance the demand for privacy against the social value of reliable population statistics. \citet{AbowdSchmutte:Privacy:AER} propose a model for determining the optimal balance between privacy and accuracy in this social choice framework. More research, both theoretical and empirical, will help data users and policy makers navigate our modern data-rich environment.

The VCG mechanism also underprovides accuracy compared with the fictitious Lindahl mechnanism. We did not identify any relation between optimal provision and the Lindahl mechanism, suggesting our results are sensitive to the setup. Our model inherits some limitations (e.g., simple market structure) of the original GR framework. We also inherit the positive aspects of their model. In particular, the results of GR's Theorems 3.1 and 3.3 are not tied to a specific auction mechanism.

In concluding, we point out possible extensions of our model. We have assumed a private data custodian must purchase privacy rights in order to use data in a published statistic. We make this assumption because it reflects the growing demand that companies be held accountable for the privacy of their customers' data through increasingly explicit means. However, if we assume companies may freely use their customers' data, then our conclusions will likely be reversed, with privacy being under-provided. To address this possibility, one might consider alternative specifications for the property rights over privacy loss.  Similarly, our model assumes that private firms are limited to the single-buyer model in their ability to elicit payment from customers for data accuracy. One might explore alternative formulations of the demand side of the market, including the use of governmental organizations as the preference aggregators in making the purchase offer. This extension mirrors the original Ghosh-Roth motivation of a researcher spending grant money to buy the statistic, then placing the result in an open-access scientific journal. Finally, our model treats the data held by a trusted data curator as ground truth. Data users concerns for accuracy will likely extend beyond our model's treatment of differential privacy as the focal source of error in the underlying database. We need better accuracy measurements and tools for incorporating other sources of error from the data generation process (e.g., edit constraints, imputation, coverage error) into the social choice problem.

%% file: appendix.tex
%!Tex root = ./PrivPubGood.tex
\subsection{Formal definition of neighboring databases}
\label{app:def_neighbors}

We can represent any database  $D$ by its un-normalized
histogram $x\in \mathbb{Z}^{\ast |\chi |}$. The notation $|\chi |$
represents the cardinality of the set $\chi $, from which database entries (rows) are drawn, and $\mathbb{Z}^{\ast }$ is
the set of non-negative integers. Each entry in $x$, $x_{i}$, is the number
of elements in the database $D$ of type $i\in \chi $. The $\ell _{1}$
norm of $x$ is
\begin{equation}
||x||_{1}=\sum_{i=1}^{|\chi |}\left\vert x_{i}\right\vert .  \label{eq:l1}
\end{equation}%
Observe that $||x||_{1}=N$, the number of records in the database. Given two
histograms, ${x}$ and ${y}$, $||{x}-{y}||_{1}$ measures the number of
records that differ between ${x}$ and ${y}$. We define \textit{adjacent
histograms} as those with equal $\ell_{1}$ norm and between which the $\ell _{1}$ distance is $2$.%

If $x$ is the histogram representation of $D$, $y$ is the histogram
representation of $D^{\prime },$ and $D^{\prime }$ is constructed from $D $
by modifying exactly one row, then $||x||_{1}=||y||_{1}$ and $||x-y||_{1}=2$. So, $D$ and $D^{\prime } $
are adjacent databases and $x$ and $y$ are the adjacent histogram
representations of $D$ and $D^{\prime }$, respectively. Some caution is
required when reviewing the related literature because definitions may be stated
in terms of adjacent databases or adjacent histograms.

\subsection{Translation of the Ghosh-Roth Model  to Our Notation}
\label{app:GR}

In this appendix we show that the results in our Section \ref%
{sec:suboptimality}, based on the definitions in the text using database
histograms and normalized queries, are equivalent to the results in %
\citet{Ghosh:Auction:GEB:2015}. In what follows, definitions and
theorems tagged GR refer to the original Ghosh and Roth (GR, hereafter)
paper. Untagged definitions and theorems refer to our results in the text.

GR model a database $D\in \left\{ 0,1\right\} ^{n}$ where there is a single
bit, $b_{i}$, taking values in $\left\{ 0,1\right\} $ for a population of
individuals $i=1,\ldots ,n$. In GR-Definition 2.1, they define a query
release mechanism $A\left( D\right) $, a randomized algorithm that maps $%
\left\{ 0,1\right\} ^{n}\rightarrow
%TCIMACRO{\U{211d} }%
%BeginExpansion
\mathbb{R}
%EndExpansion
$, as $\varepsilon _{i}$-differentially private if for all measurable
subsets $S$ of $%
%TCIMACRO{\U{211d} }%
%BeginExpansion
\mathbb{R}
%EndExpansion
$ and for any pair of databases $D$ and $D^{\left( i\right) }$ such that $%
H\left( D,D^{\left( i\right) }\right) =1$%
\begin{equation*}
\frac{\Pr \left[ A\left( D\right) \in S\right] }{\Pr \left[ A\left(
D^{\left( i\right) }\right) \in S\right] }\leq e^{\varepsilon _{i}}
\end{equation*}%
where $H\left( D,D^{\left( i\right) }\right) $ is the Hamming distance
between $D$ and $D^{\left( i\right) }$.
Notice that their use of the Hamming distance to define neighboring databases is consistent with our use of ``bounded'' differential privacy.
However, this is not the standard definition of $\varepsilon $%
-differential privacy, which they take from \citet{Dworketal:2006}, because
a \textquotedblleft worst-case\textquotedblright\ extremum is not included.
The parameter $\varepsilon _{i}$ is specific to individual $i$. The amount
of privacy loss algorithm $A$ permits for individual $i$, whose bit $b_{i}$
is the one that is toggled in $D^{\left( i\right) }$, is potentially
different from the privacy loss allowed for individual $j\neq i$, whose
privacy loss may be $\varepsilon _{j}>\varepsilon _{i}$ from the same
algorithm. In this case, individual $j$ could also achieve $\varepsilon _{j}$%
-differentially privacy if the parameter $\varepsilon _{i}$ were substituted
for $\varepsilon _{j}$. To refine this definition so that it also
corresponds to an extremum with respect to each individual, GR-Definition
2.1 adds the condition that algorithm $A$ is $\varepsilon _{i}$-\textit{%
minimally differentially private} with respect to individual $i$ if
\begin{equation*}
\varepsilon _{i}=\arg \inf_{\varepsilon }\left\{ \frac{\Pr \left[ A\left(
D\right) \in S\right] }{\Pr \left[ A\left( D^{\left( i\right) }\right) \in S%
\right] }\leq e^{\varepsilon }\right\} ,
\end{equation*}%
which means that for individual $i$, the level of differential privacy
afforded by the algorithm $A\left( D\right) $ is the smallest value of $%
\varepsilon $ for which algorithm $A$ achieves $\varepsilon $-differential
privacy for individual $i$. In GR $\varepsilon _{i}$-differentially private
always means $\varepsilon _{i}$-minimally differentially private.

GR-Fact 1, stated without proof, but see
\citet[][p. 42-43
]{Dwork:Roth:journal:version:2014} for a proof, says that $\varepsilon _{i}$%
-minimal differential privacy composes. That is, if algorithm $A\left(
D\right) $ is $\varepsilon _{i}$-minimally differentially private, $T\subset
\left\{ 1,\ldots ,n\right\} ,$ and $D,D^{\left( T\right) }\in \left\{
0,1\right\} ^{n}$ with $H\left( D,D^{\left( T\right) }\right) =\left\vert
T\right\vert $, then%
\begin{equation*}
\frac{\Pr \left[ A\left( D\right) \in S\right] }{\Pr \left[ A\left(
D^{\left( T\right) }\right) \in S\right] }\leq e^{\left\{
\sum\nolimits_{i\in T}\varepsilon _{i}\right\} },
\end{equation*}%
where $D^{\left( T\right) }$ differs from $D$ only on the indices in $T$.

In the population, the statistic of interest is an unnormalized query%
\begin{equation*}
s=\sum_{i=1}^{n}b_{i}.
\end{equation*}%
The $\varepsilon _{i}$-minimally differentially private algorithm $A\left(
D\right) $ delivers an output $\hat{s}$ that is a noisy estimate of $s$,
where the noise is induced by randomness in the query release mechanism
embedded in $A$. Each individual in the population when offered a payment $%
p_{i}>0$ in exchange for the privacy loss $\varepsilon _{i}>0$ computes an
individual privacy cost equal to $\upsilon _{i}\varepsilon _{i}$, where $%
\upsilon _{i}>0$, $p\equiv \left( p_{1},\ldots ,p_{n}\right) \in
%TCIMACRO{\U{211d} }%
%BeginExpansion
\mathbb{R}
%EndExpansion
_{+}^{n}$, and $\upsilon \equiv $ $\left( \upsilon _{1},\ldots ,\upsilon
_{n}\right) \in
%TCIMACRO{\U{211d} }%
%BeginExpansion
\mathbb{R}
%EndExpansion
_{+}^{n}$.

GR define a mechanism $M$ as a function that maps $%
%TCIMACRO{\U{211d} }%
%BeginExpansion
\mathbb{R}
%EndExpansion
_{+}^{n}\times \left\{ 0,1\right\} ^{n}\rightarrow
%TCIMACRO{\U{211d} }%
%BeginExpansion
\mathbb{R}
%EndExpansion
\times
%TCIMACRO{\U{211d} }%
%BeginExpansion
\mathbb{R}
%EndExpansion
_{+}^{n}$ using an algorithm $A\left( D\right) $ that is $\varepsilon
_{i}\left( \upsilon \right) $-minimally differentially private to deliver a
query response $\hat{s}\in
%TCIMACRO{\U{211d} }%
%BeginExpansion
\mathbb{R}
%EndExpansion
$ and a vector of payments $p\left( \upsilon \right) \in
%TCIMACRO{\U{211d} }%
%BeginExpansion
\mathbb{R}
%EndExpansion
_{+}^{n}$. GR-Definition 2.4 defines individually rational mechanisms.
GR-Definition 2.5 defines dominant-strategy truthful mechanisms. An
individually rational, dominant-strategy truthful mechanism $M$ provides
individual $i$ with utility $p_{i}\left( \upsilon \right) -\upsilon
_{i}\varepsilon _{i}\left( \upsilon \right) \geq 0$ and $p_{i}\left(
\upsilon \right) -\upsilon _{i}\varepsilon _{i}\left( \upsilon \right) \geq
p_{i}\left( \upsilon ^{\symbol{126}i},\upsilon _{i}^{\prime }\right)
-\upsilon _{i}\varepsilon _{i}\left( \upsilon ^{\symbol{126}i},\upsilon
_{i}^{\prime }\right) $ for all $\upsilon _{i}^{\prime }\in
%TCIMACRO{\U{211d} }%
%BeginExpansion
\mathbb{R}
%EndExpansion
_{+}^{n}$, where $\upsilon ^{\symbol{126}i}$ is the vector $\upsilon $ with
element $\upsilon _{i}$ removed.

GR define $k$-accuracy in GR-Definition 2.6 using
the deviation $\left\vert \hat{s}-s\right\vert $ from the output $\hat{s}$
produced by algorithm $A\left( D\right) $ using mechanism $M$ as%
\begin{equation*}
\Pr \left[ \left\vert \hat{s}-s\right\vert \geq k\right] \leq \frac{1}{3}.
\end{equation*}%
where we have reversed the direction of the inequalities and taken the
complementary probability to show that this is the unnormalized version
of our Definition \ref{def:acc} for a query sequence of length 1.
GR also define the normalized query accuracy level as $\alpha $,
which is identical to our usage in Definition \ref{def:acc}.

GR-Theorem 3.1 uses the GR definitions of $\varepsilon _{i}$-minimal
differential privacy, $k$-accuracy, and GR-Fact 1
composition to establish that any differentially private mechanism $M$ that
is $\left( \frac{\alpha n}{4}\right) $-accurate must purchase
privacy loss of at least $\varepsilon _{i}\geq \frac{1}{\alpha n}$ from at
least $H\geq \left( 1-\alpha \right) n$ individuals in the population.
GR-Theorem 3.3 establishes the existence of a differentially private
mechanism that is $\left( \frac{1}{2}+\ln 3\right) \alpha n$-accurate and
selects a set of individuals $H\subset \left\{ 1,\ldots ,n\right\} $ with $%
\varepsilon _{i}=\frac{1}{\alpha n}$ for all $i\in H$ and $\left\vert
H\right\vert =\left( 1-\alpha \right) n$.

In order to understand the implications of GR-Theorems 3.1 and 3.3 and our
arguments about the public-good properties of differential privacy, consider
the application of GR-Definition 2.3 ($\func{Lap}\left( \sigma \right) $
noise addition) to construct an $\varepsilon $-differentially private
response to the counting query based on GR-Theorem 3.3 with $\left\vert
H\right\vert =\left( 1-\alpha \right) n$ and the indices ordered such that $%
H=\left\{ 1,\ldots ,\left\vert H\right\vert \right\} $.
% Assume, as we do in
% Theorem \ref{theorem:suboptimality} and as GR do in their proof of
% GR-Theorem 3.3, that $n$ is sufficiently large that we can ignore the
% difference between $\left( 1-\alpha \right) n$ and $\func{ceil}\left( \left(
% 1-\alpha \right) n\right) $.
The resulting answer from the query response
mechanism is%
\begin{equation*}
\hat{s}= \sum_{i=1}^{H}b_{i}+\frac{\alpha n}{2} + \func{Lap}\left( \frac{1}{\varepsilon }\right) ,
\end{equation*}%
which is the counting query version of equation $\left( \ref{eqn:GRquery}\right) $ in the text. Note the bias correction term $\alpha n/2$ is adjusted in equation $\left( \ref{eqn:GRquery}\right)$ as necessitated by our use of $\left(\alpha, \frac{1}{3}\right)$-accuracy.    Because of
GR-Theorem 3.3, we can use a common $\varepsilon =\frac{1}{\alpha n}$ in
equation $\left( \ref{eqn:GRquery}\right) $.

If this were not true, then we would have to consider query release
mechanisms that had different values of $\varepsilon $ for each individual
in the population and therefore the value that enters equation $\left( \ref%
{eqn:GRquery}\right) $ would be much more complicated. To ensure that each
individual in $H$ received $\varepsilon _{i}$-minimally differential
privacy, the algorithm would have to use the smallest $\varepsilon _{i}$
that was produced for any individual. In addition, the FairQuery and MinCostAuction
algorithms described next would not work because they depend upon being able
to order the cost functions $\upsilon _{i}\varepsilon _{i}$ by $\upsilon _{i}
$, which is not possible unless $\varepsilon _{i}$ is a constant or $%
\upsilon _{i}$ and $\varepsilon _{i}$ are perfectly positively correlated.
Effectively, GR-Theorem 3.3 proves that achieving $\left(\alpha ,\beta
\right) $-accuracy with $\varepsilon $-differential privacy requires a
mechanism in which everyone who sells a data-use right gets the best
protection (minimum $\varepsilon _{i}$ over all $i\in H$) offered to anyone
in the analysis sample. If a change in the algorithm's parameters results in a
lower minimum $\varepsilon _{i}$, everyone who opts to use the new parameterization
receives this improvement. In addition, we argue in the text that when such
mechanisms are used by a government agency they are also non-excludable
because exclusion from the database violates equal protection provisions of
the laws that govern these agencies.

Next, GR analyze algorithms that achieve $O\left( an\right) $-accuracy by
purchasing exactly $\frac{1}{\alpha n}$ units of privacy loss from exactly $%
\left( 1-\alpha \right) n$ individuals. Their algorithms \textit{FairQuery}
and \textit{MinCostAuction} have the same basic structure:

\begin{itemize}
\item Sort the individuals in increasing order of their privacy cost, $%
\upsilon _{1}\leq \upsilon _{2}\leq \ldots \leq \upsilon _{n}$.

\item Find the cut-off value $\upsilon _{k}$ that either exhausts a budget
constraint (FairQuery) or meets an accuracy constraint (MinCostAuction).

\item Assign the set $H=\left\{ 1,\ldots ,k\right\} .$

\item Calculate the statistic $\hat{s}$ using a differentially private
algorithm that adds Laplace noise with just enough dispersion to achieve the
required differential privacy for the privacy loss
purchased from the members of $H$.

\item Pay all members of $H$ the same amount, a function of $\upsilon _{k+1}$%
; pay all others nothing.
\end{itemize}

To complete the summary of GR, we note that GR-Theorem 4.1 establishes that
FairQuery is dominant-strategy truthful and individually rational.
GR-Proposition 4.4 establishes that FairQuery maximizes accuracy for a given
total privacy purchase budget in the class of all dominant-strategy
truthful, individually rational, envy-free, fixed-purchase mechanisms.
GR-Proposition 4.5 proves that their algorithm MinCostAuction is a VCG
mechanism that is dominant-strategy truthful, individually rational and $%
O\left( \alpha n\right) $-accurate. GR-Theorem 4.6 provides a lower bound on
the total cost of purchasing $k$ units of privacy of $k\upsilon _{k+1}$ \
GR-Theorem 5.1 establishes that for $\upsilon \in
%TCIMACRO{\U{211d} }%
%BeginExpansion
\mathbb{R}
%EndExpansion
_{+}^{n}$, no individually rational mechanism can protect the privacy of
valuations $\upsilon $ with $\left( k,\beta\right) $-accuracy for $k<%
\frac{n}{2}$.

In our application of GR, we use $N$ as the total population. Our $\gamma
_{i}$ is identical to the GR $\upsilon _{i}$. We define the query as a
normalized query, which means that query accuracy is defined in terms of $%
\alpha $ instead of $k$; hence, our implementation of the VCG mechanism
achieves $\left( \alpha ,\beta\right)$ where the inclusion of $\beta$ generalizes GR's implicit restriction to $\beta=\frac{1}{3}$ in their accuracy definition. We define the individual amount of privacy loss in the same manner as GR.

% Local Variables:
% mode: latex
% TeX-master: abowd-schmutte-privacy
% End: